\documentclass{elsart}
\usepackage{graphicx,harvard,amssymb}
\journal{New Astronomy}

\makeatletter
\def\elsartstyle{%
	\def\normalsize{\@setfontsize\normalsize\@xiipt{14.5}}
	\def\small{\@setfontsize\small\@xipt{13.6}}
	\let\footnotesize=\small
	\def\large{\@setfontsize\large\@xivpt{18}}
	\def\Large{\@setfontsize\Large\@xviipt{22}}
	\skip\@mpfootins = 18\p@ \@plus 2\p@
	\normalsize
}
\makeatother

\def\url#1{{\ttfamily\def\/{/\discretionary{}{}{}}#1}}

\def\mincir{\raise -2.truept\hbox{\rlap{\hbox{$\sim$}}\raise5.truept
\hbox{$<$}\ }}
\def\magcir{\raise -4.truept\hbox{\rlap{\hbox{$\sim$}}\raise5.truept
\hbox{$>$}\ }}

\begin{document}

\begin{frontmatter}

\title{Cosmic opacity to CMB photons and polarization measurements}

\author{L.P.L. Colombo\thanksref{lc}} \& 
\author{S.A. Bonometto\thanksref{sb}}

\address{Physics Department G. Occhialini, Universit\`a degli Studi di
Milano--Bicocca, Piazza della Scienza 3, I20126 Milano (Italy)}
\address{I.N.F.N., Via Celoria 16, I20133 Milano (Italy)}

\thanks[lc]{E-mail: loris.colombo@mib.infn.it}
\thanks[sb]{E-mail: silvio.bonometto@mib.infn.it}

\begin{abstract}
Anisotropy data analysis leaves a significant degeneracy between primeval 
spectral index ($n_s$) and cosmic opacity to CMB photons ($\tau$).
Low--$l$ polarization measures, in principle, can remove it.
We perform a likelihood analysis to see how cosmic variance possibly
affects such a problem. We find that, for
a sufficiently low noise level ($\sigma^P_{pix}$) and if 
$\tau$ is not negligibly low, the degeneracy is greatly reduced, while the
residual impact of cosmic variance on $n_s$ and $\tau$ determinations
is under control. On the contrary, if $\sigma^P_{pix}$ is too high, 
cosmic variance effects appear to be magnified. We apply general
results to specific experiments and find that, if favorable 
conditions occur, it is possible
that a 2--$\sigma$ detection of a lower limit on $\tau$ is
provided by the SPOrt experiment. Furthermore, if the PLANCK
experiment will measure polarization with the expected precision,
the error on low--$l$ harmonics is adequate to determine $\tau$,
without significant magnification of the cosmic variance.
This however indicates that high sensitivity might be more important 
than high resolution in $\tau$ determinations.
We also outline that a determination of $\tau$ is critical to
perform detailed analyses on the nature of dark energy and/or
on the presence of primeval gravitational waves.

\end{abstract}

\begin{keyword}
Cosmic Microwave Background \sep
Cosmology: Cosmological Parameters 
\PACS 98.70.Vc \sep 98.80.-k
\end{keyword}
\end{frontmatter}

\section{Introduction}

Data on cosmic microwave background (CMB) anisotropy and polarization
can provide effective constraints on cosmological parameters. However,
while significant anisotropy observations are available (see, e.g.,
Smoot et al. 1992; de Bernardis et al. 2000; Hanany et al. 2000; 
Halverson et al. 2002; Sievers et al. 2002; Scott et al. 2002) and further
anisotropy data are expected from experiments in progress, detailed
data on the weaker polarization signals may become available only 
within a few years.
In turn, dark energy (DE) parameters or the amplitude of primeval 
gravitational waves (GW) are just mildly constrained by anisotropy data,
while polarization would significantly depend on them. 
A peculiar situation then holds for the cosmic optical depth to 
CMB photons ($\tau$), due to reionization. Anisotropy data only mildly 
constrain it, while polarization data would provide more stringent
constraints; of course, greater $\tau$ values would ease its detection, 
but, in turn, greater $\tau$ values would also ease the determination 
of DE and/or GW parameters.

This paper tries to investigate which polarization experiment(s) can 
determine $\tau$. The success of an experiment, however, will also 
depend on the value of $\tau$ itself. In particular, we shall consider 
the SPOrt experiment (Macculi et al. 2000; Carretti et al. 2000; 
Peverini et al. 2001), planned to perform a (nearly) full--sky 
polarization measure aboard of the ISS in 2004, with an angular 
resolution similar to COBE.

When only anisotropy data are considered, there is a degeneracy 
between the effects of varying $\tau$ or the primeval spectral
index $n_s$ (Jungman et al. 1996; 
Zaldarriaga, Spergel \& Seljak 1997; Eisenstein, Hu \&
Tegmark 1999; Efstathiou \& Bond 1999). 
The most stringent limits on $\tau$, based on anisotropy 
data, are provided by Stompor et al. (2001), who used the outputs of 
recent balloon experiments to set $\tau\lesssim 0.4$ at the 2--$\sigma$ 
level, once $n_s$ is assumed $\le1.2$. (Such limits will be included in a 
number of figures here below and denominated {\it Stompor limits}). 
Full--sky anisotropy measures from space, like 
MAP\footnote{\url{http://map.gsfc.nasa.gov}} or 
PLANCK\footnote{\url{http://astro.estec.esa.nl/Planck}}, might 
improve the {\it Stompor limits} by a factor $\sim 2$.
In turn, a residual uncertainty ($\sim 0.05$) on the value of $n_s$ is 
related to our scarce knowledge of $\tau$. Of course, a more precise
determination of $n_s$ would really be welcome, first of all
to shed new light on the mechanism generating primeval fluctuations.
Furthermore, within a generic inflationary model, only for $n_s < 1$
primeval GW may be expected to exist.

The impact that a variation of $\tau$ may have on CMB polarization
can be seen from Fig.~1; here, for a spatially flat cosmological model
(density parameters: $\Omega_m = 0.35$, $\Omega_b = 0.05$; Hubble 
constant in units of 100 km/s/Mpc: $h=0.65$; $n_s = 1$; dark energy 
is an ordinary cosmological constant) and for different values 
of $\tau$, we plot the dependence on $l$ of the angular spectra 
$C_l^{A}$ with $A=T,~E,~B,~X$, standing for anisotropy, 
$E$-- and $B$--mode polarization and anisotropy--polarization 
cross correlation respectively; 
only $E$--mode polarization is considered through this paper.
Fig.~1 shows that the main dependence on $\tau$ actually occurs 
at low $l$ and for the polarization and cross--correlation spectra; 
comparatively, the $\tau$ dependence of spectra, at $l 
\gtrsim 30$, is milder. This point is important  
to devise the right polarization experiment, which should aim to
great sensitivity, while, for $\tau$ determination, high resolution
data might not bear a direct relevance.
Accordingly, the angular resolution of the SPOrt experiment does not
prevent it from achieving cosmologically significant results.

Here, as well as in most recent literature, we assume that
the whole cosmic opacity arises from an almost complete ionization of hydrogen,
after a suitable redshift. More complex reionization histories
may have to be taken into account to discuss future data (Bruscoli,
Ferrara \& Scannapieco 2002; Venkatesan 2002), but
would not modify general conclusions.

For low--$l$ spectral components, varying $n_s$ has a milder effect 
than varying $\tau$. If the whole angular spectrum is considered, 
other parameters, like $\Omega_{tot}$, $\Omega_m$, $\Omega_b$ (total,
matter and baryon density parameters), may be determined independently 
from low--$l$ features. Here we shall assume that they are
already known from such high--$l$ anisotropy measures.

Dealing with low $l$'s, a significant part of our analysis will 
be devoted to inspect cosmic variance. Its impact on $C_l$ determination
is known. Here, however, we are interested in measuring different 
quantities and a knowledge of $C_l$'s is not even required. We 
shall explore this point by simulating and analyzing a large number 
of artificial data sets. Among other things, we shall see that
the SPOrt experiment, although far from granting a stringent
$\tau$ determination, may improve the {\it Stompor limit} in a
significant number of cases.

In the next section we shall first debate which range of values
can be expected for $\tau$. We shall then describe how artificial 
data sets are produced. In section 4 we shall show the results of 
their analysis, in a number of different experimental conditions and 
considering different $\tau$ values. In section 5, conclusions will be drawn.

\section{The cosmic opacity to CMB photons}
In the present cosmic epoch, although diffuse baryonic materials are 
almost completely ionized, the scattering time for CMB photons $t_s \simeq 
4.45 \cdot 10^{18} \Omega_b^{-1} h^{-2}$s ($h$: Hubble parameter in 
units of 100~km/s/Mpc) exceeds the Hubble time $t_H = 3.09 \cdot 10^{17} 
h^{-1}$s and the Universe is transparent to CMB photons. These figures 
can be extrapolated to the past, but, while $t_s \propto a^3$ ($a$: the 
scale factor), the dependence of $t_H$ on $a$ depends on the cosmological 
model. In an expansion regime dominated by non--relativistic matter, a 
fully ionized Universe is opaque to CMB at $z > z_{op} \simeq 5\, 
(\Omega_b h)^{-2/3}$. Therefore, assuming $\Omega_b h^2 \leq 0.022$
(standard BBNS limits; see, e.g., Dolgov 2002 for a recent review) 
and $h \simeq 0.5$, we have $z_{op} > 64$, 
safely above the expected reionization redshift $z_{ri}$. No reasonable 
cosmological model allows a substantial reduction of such $z_{op}$.

On the contrary, the precise value of $\tau (\ll 1)$, does depend on the 
model and, in particular, on $z_{ri}$. Recent data on high--$z$ QSO's 
(Djorgovski et al. 2001; Becker et al. 2001) indicate that at least a 
fraction of neutral hydrogen is present, in the intergalactic medium, 
at $z \gtrsim 6$. These very authors, however, warn us against
an immediate conclusion that $z_{ri} \simeq 6$. Reionization can be
a slow and/or patchy process and our present knowledge of primeval object 
formation can be consistent with an effective $z_{ri} \simeq 6$ or a 
significantly greater value. In Fig.~2 we show how $\tau$ is related 
to $z_{ri}$ and model features, considering a wide set of cosmological 
models. For $z_{ri} \lesssim 15$, values of $\tau$ up to $\sim 0.15$--0.20 
are licit.

\section{The artificial data sets}
Artificial data sets will be built with reference to the features
of the Sky Polarization Observatory (SPOrt) experiment, selected 
by the European Space Agency (ESA) to fly on the International Space 
Station (ISS) in 2004, aiming to measure the linear polarization 
of the diffuse sky radiation at 22, 32 and 90 GHz, using a set of 
polarimeters fed by corrugated horns with an angular resolution 
(FWHM) of $7^o$. At variance from other space experiments (MAP 
and PLANCK), SPOrt has been explicitly designed to measure
the Stokes parameters Q and U and to minimize systematics and 
instrumental polarization. SPOrt will observe a sky area with declination 
$|\delta | \le 51.6^o$ ($\sim 80\%$ of the whole sky). The lifetime of 
the experiment is at least 1.5 years, but an extension is not unlikely; 
data will be binned in a number of pixels ranging from 662 to several 
thousands, depending on the effective duration of the flight and 
efficiency. 

In this analysis we shall assume that pixels are distributed according 
to the HEALPix\footnote{\url{http://www.eso.org/science/healpix/}} 
package, with $N^{side} = 8$ or 16, and smoothed with a 
Gaussian beam of FWHM of $7^o$. The two choices of $N^{side}$ allow
to confirm that the best exploitation of the observational output
is achieved when the average angular distance between pixel centers
is $\sim$ half FWHM. For the low (high) resolution cases we shall
therefore have, on the whole sky, 768 (3072) pixels, whose centers 
lie at an average angular distance of $\sim 7.3^o$ ($3.7^o$). Once 
polar caps are excluded, there remain 600 (2448) pixels, providing 
measures of the Stokes parameters $Q$ and $U$.

SPOrt polarimeters will provide no anisotropy data. Such information 
will be however available through other experiments (COBE, MAP) and
we shall assume here that no peculiar problem arises in correlating
SPOrt and anisotropy data. Of course, anisotropy data probe a
different sky area, i.e. the whole sky, excluding the area
with declination $| \delta | < 20^o$, where galactic contamination 
is severe. (On the contrary, such contamination can be assumed 
to be under control, for polarization data; see, e.g., Bruscoli et
al. 2002; Tucci et al. 2002.)
For the coarse (fine) pixelization cases, as above, in this sky area 
there are 480 (1984) pixels. For 376 (1360) of them, both anisotropy 
and polarization data are supposed to be available.

Random noise ($\sigma^P_{pix}$) will be included in artificial data, 
assuming it
to be uncorrelated, both among pixels and between $Q$ and $U$, 
in the case of polarization measures. We shall assume $\sigma^T_{pix}
= 1\, \mu$K ($2 \mu$K)
for temperature data (as expected for MAP measurements scaled to the
SPOrt resolution). On the contrary, one of the main variables we 
consider is the noise level for polarization data.

SPOrt data will be characterized by 
$\sigma^P_{pix} \simeq 5.3\,{N^{side} \over 8}\, \sqrt{{1.5 \over
\lambda} {0.5 \over \epsilon}}\, \mu$K per pixel, where $\lambda$ is the
experiment's lifetime in years while $\epsilon$ indicates the detection
efficiency. Our analysis will
consider $\sigma^P_{pix}$'s in the interval comprised from 0.5 to 5$\, \mu$K 
per pixel, with coarse pixelization, corresponding to 1 to 10$\, \mu$K 
per pixel, with fine pixelization.

The expected angular spectra $C_l^{T,E,B,X}$, were produced with the
program CMBFAST
\footnote{\url{http://physics.nyu.edu/matiasz/CMBFAST/cmbfast.html}}, 
starting from a spatially flat cosmological model
with $\Omega_m = 0.35$, $\Omega_b = 0.05$, $h=0.65$. Dark energy
is an ordinary cosmological constant. Several values of $n_s$ were 
considered, but we shall report results only for $n_s = 1$.
Other close $n_s$ values yield quantitatively similar results and
the potentiality of the experiment does not depend on $n_s$, in any
appreciable way. On the contrary, our analysis aimed to exploring
$\tau$ values ranging from 0.035 to 0.25. Cosmic variance was taken
into account by considering 1000 independent realizations for each
$\tau$ and $\sigma^P_{pix}$ pair.

\section{Data analysis}
The first aim of our analysis amounts to finding the likelihood distribution
on the plane $\tau$--$n_s$. The number of pixels for anisotropy and 
polarization ($N_T$ and $N_P$) in our (artificial) data are different. 
Let then $T_j$ be the anisotropy measured in $N_T$ pixels and $Q_j$ 
and $U_j$ the Stokes parameters measured in $N_P$ pixels. 
In general, let us define vectors ${\bf x} \equiv (T_1,.....,T_{N_T}, 
Q_1,.....,Q_{N_P}, U_1,.....,U_{N_P})$, of $N_s = N_T + 2 N_P$ components, 
defining an observed state of anisotropy and polarization. Once
a pair of values $\tau$--$n_s$ is assigned, the angular spectra
$C_{l}^A= ( C_{l}^T,C_{l}^E,C_{l}^X )$ are univocally determined.
On the contrary, a data vector ${\bf d}$, of $N_s$ components, built 
from them, is just a {\it realization} of such model:
once the $N_s$ component vector ${\bf d}$ is assigned, the pair
of values $\tau$--$n_s$ is not univocally fixed.

A function
$$
{L}({\bf d}|C^A_l) \propto [\, \det {\bf M}\, ]^{-{1 \over 2}} \exp 
\big[ -{1 \over 2}{\bf d^T} {\bf M}^{-1} {\bf d} \big]
\eqno (1)
$$
shall then be built, to yield the likelihood of a given set of $C^A_l$ 
(i.e., of a pair of $\tau$--$n_s$ values), if ${\bf d}$ is observed.
The main ingredient of $L$ is the correlation matrix 
${\bf M_{ij}} = \langle {\bf x}^T_i {\bf x}_j \rangle = {\bf S}_{ij} 
+ {\bf N}_{ij}$; here ${\bf S}_{ij} $ is the signal term and
${\bf N}_{ij}$ is due to the noise. The components ${\bf M_{ij}}$
yield the correlation between the $i$th and $j$th elements of 
data vectors ${\bf x}$ corresponding to particular choices of $C_{l}^A$,
i.e. of  $\tau$--$n_s$ values. The construction of the 
(model dependent) signal term, however, does not require to build explicitly
the vectors ${\bf x}$. The procedure to be followed, in the case when
both anisotropy and polarization data are available, is explicitly
reported by Zaldarriaga (1998).
The construction of the noise term is simpler, as we expect no
noise correlation, and the matrix ${\bf N}_{ij} =
\delta_{ij}\,{\sigma_{T,pix}}^2$ (for $i = 1,...,N_T$) and 
${\bf N}_{ij} = \delta_{ij}\,{\sigma_{P,pix}}^2$ (for $i =
N_T+1,...,N_s$) is diagonal.

In what follows, the technical role of the $C_l^A$ spectra will not
need to be further outlined and the likelihood function will be
explicitly considered to depend on $\tau$ and $n_s$.
In particular, we assume that the most probable $\tau$--$n_s$ pair, 
for a given anisotropy--polarization state ${\bf d}$, 
is the one which maximizes 
the likelihood. On the $\tau$--$n_s$ plane, we can then define
curves connecting points where $L$ corresponds to a given
fraction of its top value. The area enclosed by each such curve
corresponds to a given confidence level. In the figures below,
such fractions were selected in order that the areas enclosed
by the curves correspond to 68$\, \%$, 90$\, \%$ and 99$\, \%$
confidence levels. For the sake of simplicity, we shall refer to
them as 1--2--3$\, \sigma$ levels.

Because of the role played by low--$l$ harmonics in our analysis,
as above outlined, cosmic variance may be quite significant.
Henceforth, 1000 realizations of each model will be 
considered; 1--2--3$\, \sigma$ curves will be obtained by averaging among
the results of such realizations, so obtaining an
{\it average likelihood distribution}. Of course, this is not sufficient
to evaluate the expected result of a given experiment. We shall
therefore implement average results with histograms, describing
the distribution of results in various realizations,
suitably defined for the various cases.

\section{Results}
Because of the large number of realizations required, we considered
likelihood distributions both for coarse and fine pixelization,
corresponding to distances between pixel centers of  $\sim 7.3^o$ or $3.7^o$
(see section 3). A comparison of such distributions, for the cases
we treated in both ways, confirms that, in general, a finer pixelization
allows a better exploitation of data. Finding some general
trends, however, does not require fine pixelization analysis.

In Fig.~3, we report the likelihood distributions for $\tau=0.05$ and 
4 levels of $\sigma^P_{pix}$ (for coarse pixelization). In Fig.~4, we 
report the likelihood distribution for the same $\tau$ value and 
$\sigma^P_{pix} = 2\, \mu$K (for fine pixelization, corresponding 
to $1\, \mu$K with
coarse pixelization). Besides of the solid curves obtained by considering
both anisotropy and polarization, dashed curves are shown, obtained
from polarization data only. Stompor limits are explicitly shown
in Fig.~4, while in Fig.~3 we report Stompor's 95$\, \%$ confidence 
limit only. See the caption of Fig.~4 for further details on
Stompor limits.

From these figures it is clearly visible that upper limits
on $\tau$ can be significantly improved, in respect to previous
analyses, for $\sigma^P_{pix} \lesssim 2$--3$\, \mu$K. This is true
even if only polarization data are used. The main improvement
obtained, when using anisotropy data as well, concerns $n_s$,
for which both lower and upper limits are obtained.
When passing from coarse to fine pixelization,
$n_s$ is even better constrained. If $\sigma^P_{pix}$ is at the lowest
levels considered, $n_s$ determination is better than in Stompor's
case. 

For what concerns lower limits on $\tau$, the situation is
more critical. Detecting a lower limit, however, should be the
main aim of such kind of experiments. Fig.~4 shows that a lower
limit, in average, can be detected at the 1--$\sigma$ confidence level
at most. This is a case, however, when average estimates themselves are
insufficient. In Fig.~5, therefore, we show the distribution
of 1--$\sigma$ lower limits found, with or without taking into
account anisotropy, with fine pixelization. In the former
case, we see that, in $\sim 78\, \%$ of cases a lower limit is
found. Such percentage decreases to $\sim 70\, \%$, if no
anisotropy data could be considered. For the sake of comparison,
let us report that the above figures decrease to $\sim 76\, \%$ and
$\sim 64\, \%$, with coarse pixelization.

Going below a noise level of $2\, \mu$K per pixel (with fine
pixelization), although highly problematic from an experimental
point of view, would be however decisive. This is shown by
Fig.~6, where we see that, for $\sigma^P_{pix} = 1\, \mu$K, 
even the average 3--$\sigma$ curve indicates a detection of 
$\tau > 0$. If the distribution of lower limits in various
realizations is considered, we find that, only in $\sim 2\, \%$
of cases, no 3--$\sigma$ lower limits would be detectable.

Let us now consider the cases with greater $\tau$ values.
They are reported in Fig.~7 and 8, for $\tau = 0.10$ and 0.15.
These figures are all obtained with coarse pixelization.
They confirm that significant improvements of Stompor
upper limits on $\tau$ are obtainable for $\sigma^P_{pix} \lesssim 
2$--3$\, \mu$K.
Moreover, lower limits become safer than in the $\tau = 0.05$
case. Even with $\sigma^P_{pix} = 2\, \mu$K, if $\tau \sim 0.15$,
a 1--$\sigma$ lower limit is likely to be obtained and
a 2--$\sigma$ lower limit might be obtained as well.
This is an important point, as such noise levels are not
far from those obtainable, in the best possible conditions,
by the SPOrt experiment.

It is also significant to go above average values, for lower
limits, in all previous cases. In Fig.~9 to 11, we report the
fractions of realizations, for which lower limits were detected,
both using anisotropy--polarization  data and polarization only.
Among various possible comments, these
plots confirm that, in order to detect $\tau$ (unless $\gtrsim 0.15$),
the noise level of the SPOrt experiment is too high.
It may also be significant to compare these figures with the
expected noise level in the PLANCK measurements of polarization.
Weighting the various channels in order to minimize variance,
it is expected that $\sigma^P_{pix} \sim 9\, \mu$K, for a
FWHM of $7'$ (see, e.g., Hu \& Okamoto 2002). 
If a scaling $\sigma^P_{pix} \propto {\rm angle}^{-1}$
is assumed, this yields $\sigma^P_{pix} \sim 0.4\, \mu$K, if reported
to our fine pixelization. At such noise level, no magnification of
the intrinsic cosmic variance seems to occur.

\section{Conclusions}
In this paper we performed a likelihood analysis, to
determine how a polarization experiment could break
the degeneracy between $n_s$ and $\tau$ determinations,
found in pure anisotropy analyses.
To this aim we built artificial data, consistent with
an experiment in progress to detect polarization,
but pushing noise even well below the expected experimental
level. The experiment considered will cover 80$\, \%$ of
the whole sky and artificial data were built assuming such
sky coverage. If data were available for the full
sky, a slight improvement of the signal can be expected. 
Artificial data were then analyzed. In respect to future 
observers, we have however the critical advantage to
be able to analyze as many skies (realizations) we need.
This is significant, in our case, as the $n_s$--$\tau$
degeneracy, in principle, is readily overcome from low--$l$
spectral components, where, however, cosmic variance can be critical.
Our results, however, indicate that cosmic variance
is not such a severe limitation, if a sufficiently low noise level is attained.
This is one of the most important conclusions of our analysis,
as it confirms that, to detect $\tau$, polarization experiments should aim to
high sensitivity, while high resolution does not bear a direct relevance.

The critical issue, in this respect, is whether and when firm lower
limits on $\tau$ are obtainable. The noise level to be
attained, to implement such aim, obviously depends on
the physical value of $\tau$ itself. Available constraints
on $\tau$, besides that from CMB anisotropy data, can be also
related to the detection of hydrogen lines in high--$z$
QSO spectra. This however leaves us a still significant range
of possible cosmic opacities to CMB photons. 

If $\tau$ is not negligibly low, the noise level below which
we expect that lower limits on $\tau$ can be obtained is
$\sigma^P_{\rm pix} \simeq 1$--2$\, \mu$K, for an instrument
with an angular resolution similar to COBE. A detection, at
higher $\sigma^P_{\rm pix} $, cannot be however excluded.
We compared such general conclusion with the 
actual expected features of the SPOrt experiment and found that, 
for some of the greatest values for $\tau$ allowed by the cosmic
reionization physics, and assuming the best possible performance 
of the experiment, a detection of $\tau$, at the 2--$\sigma$
level, is possible.

Acknowledgements -- The public programs CMBFAST, by U. Seljak \&
M. Zaldarriaga, was widely used here, together with its generalization 
to dynamical dark energy models due to R. Mainini. The public program 
HEALPix, by K.M. G\`orski et al. was also widely used in the
preparation of this work. Thanks are also due to Stefano Cecchini and 
Marco Tucci for discussions and comments.

\newpage

\begin{figure}
\begin{center}
\includegraphics*[width=9cm]{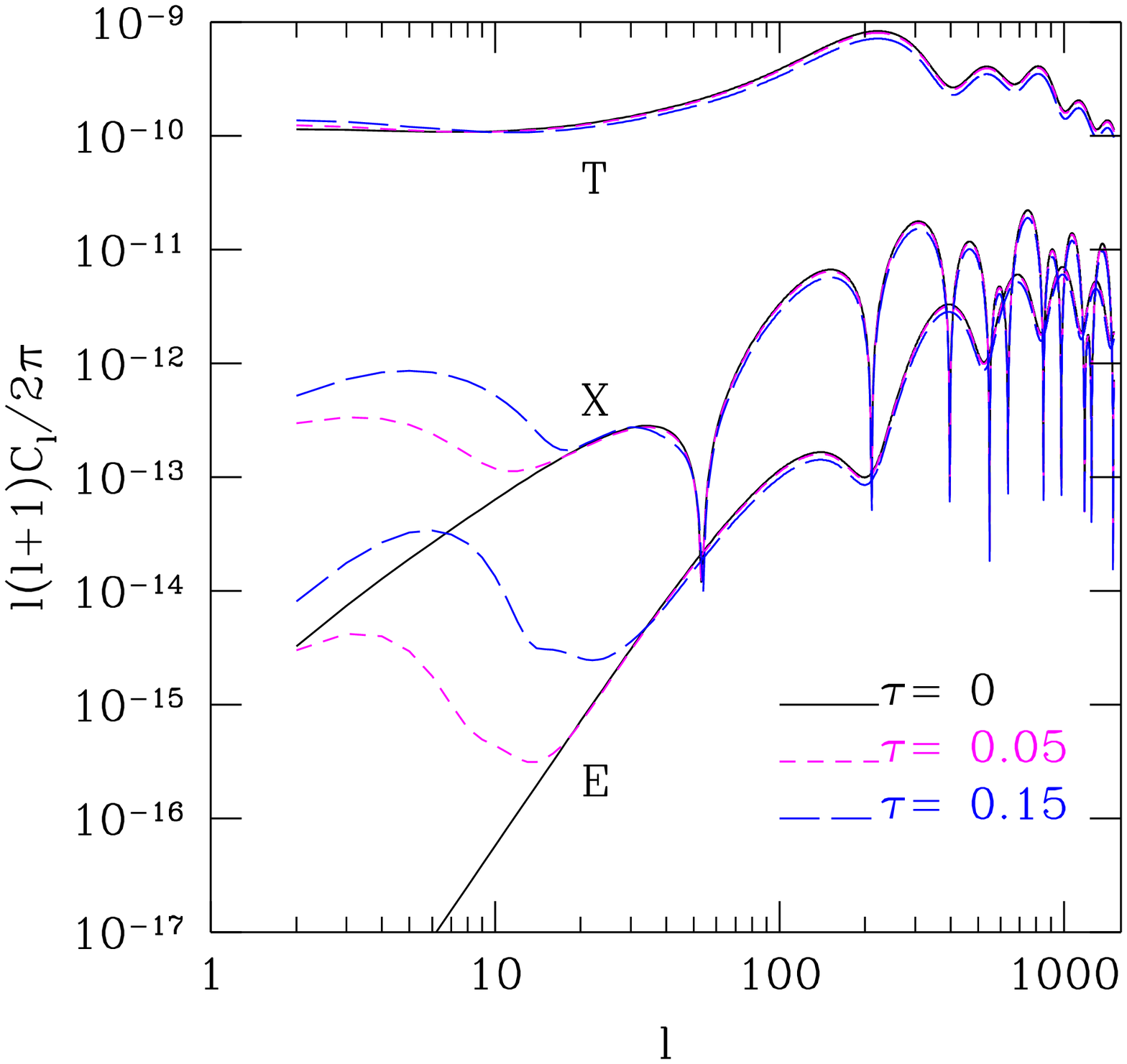}
\end{center}
\caption{Angular spectra with different $\tau$ values.}
\label{1fig}
\end{figure}

\begin{figure}
\begin{center}
\includegraphics*[width=9cm]{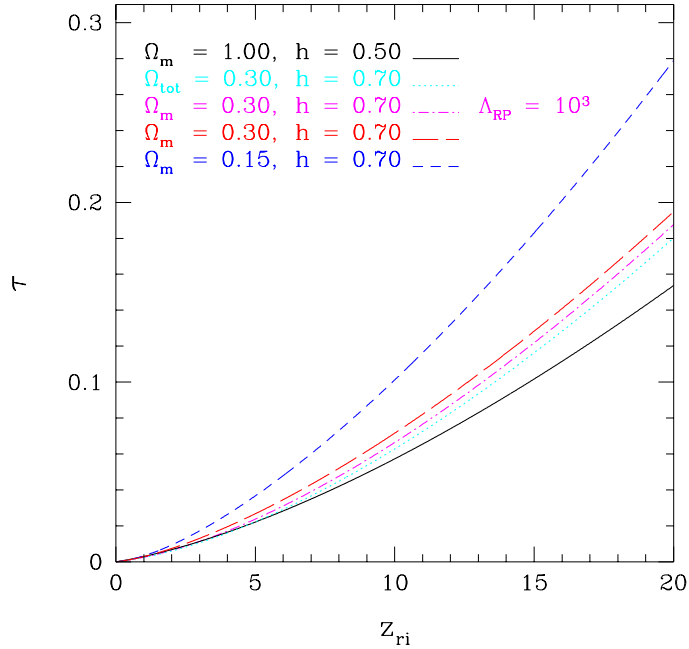}
\end{center}
\caption{Dependence of $\tau$ on the reionization redshift $z_{ri}$ 
in various cosmological models.
The only model with total density parameter $\Omega_{\rm tot} \neq 1$ 
is the second one. In the third model, dark energy
is due to a scalar field, self--interacting through a Ratra--Peebles
potential (1988), with $\Lambda_{RP}/{\rm GeV} = 10^3$. In all models 
$\Omega_b h^2 = 0.022$. We assume (almost) all
baryons to be ionized after $z_{ri}$.}
\label{2fig}
\end{figure}

\begin{figure}
\begin{center}
\includegraphics*[width=9cm]{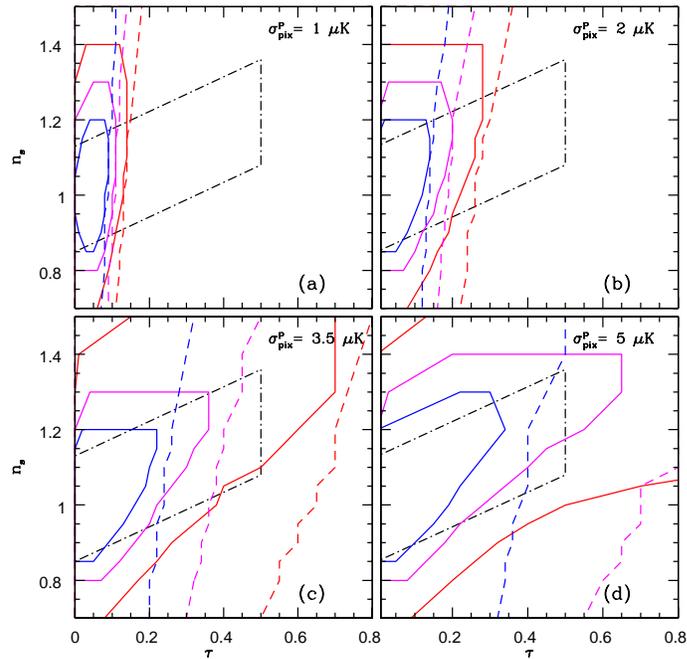}
\end{center}
\caption{Average likelihood distributions, in a model with $\tau = 0.05$,
if the pixel--pixel angular distance is 7.3$^o$. Solid curves
are obtained considering both anisotropy and polarization data;
dashed curves are obtained with polarization data only; the
corresponding confidence levels are reported in the text.}
\label{3fig}
\end{figure}

\begin{figure}
\begin{center}
\includegraphics*[width=9cm]{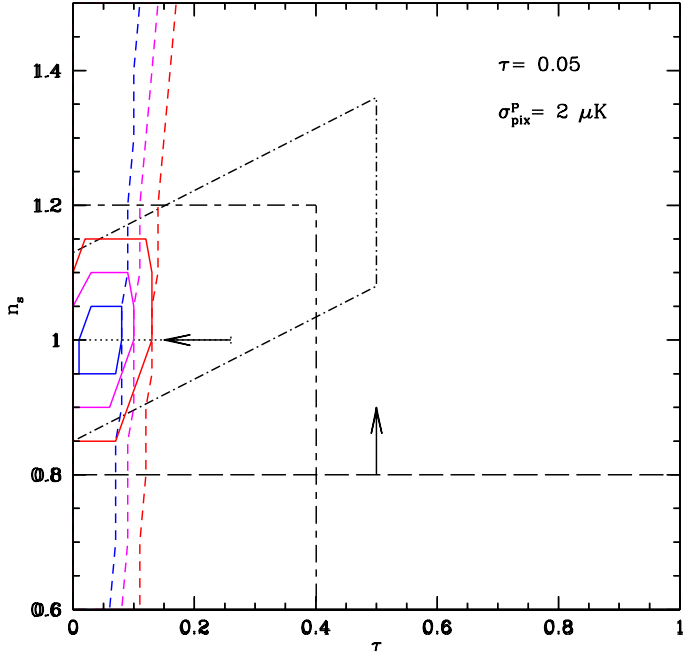}
\end{center}
\caption{Average likelihood distributions, in a model with 
$\tau = 0.05$ and $\sigma^P_{pix} = 2\, \mu$K, if the pixel--pixel 
angular distance is 3.7$^o$. Solid and dashed curves are as in Fig.~3.
The limits found by Stompor at al (2001), using anisotropy data
from balloon experiments, are also shown: The dot--dashed contour
limits the 95$\, \%$ confidence area, without any prior (this
contour is reported in other plots as well). Assuming that
$n_s < 1.2$, they set the limit $\tau < 0.4$, at the 95$\, \%$ c.l., 
(long--short dashed contour). With the prior $n_s \equiv 1$, instead,
$\tau < 0.26$, as shown by the arrow followed by a dotted line.
Without any prior, however, they find that, at the 99$\, \%$ c.l.,
$n_s > 0.8$ (long dashed line).}
\label{4fig}
\end{figure}

\begin{figure}
\begin{center}
\includegraphics*[width=9cm]{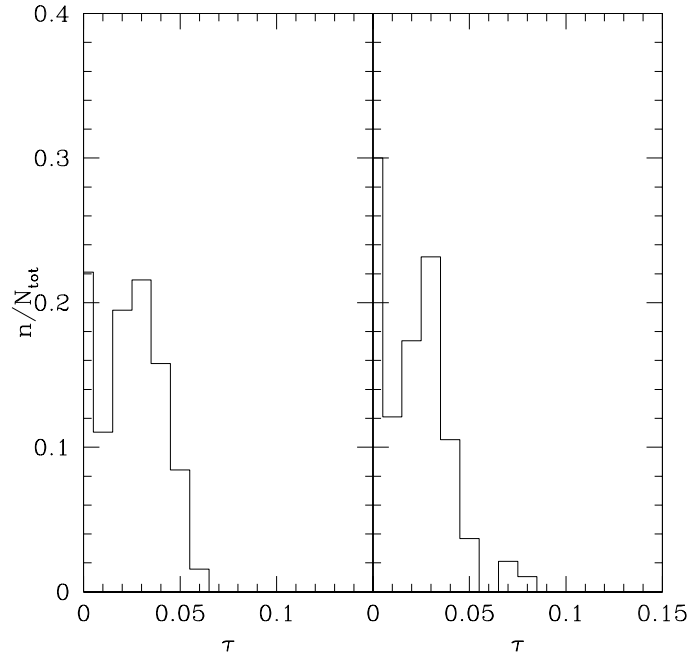}
\end{center}
\caption{Distribution of 1--$\sigma$ lower limits, for $\tau=0.05$ and
$\sigma^P_{pix} = 2\, \mu$K, if the pixel--pixel angular distance is 3.7$^o$.
The l.h.s. (r.h.s.) plot includes (excludes) anisotropy data.}
\label{5fig}
\end{figure}

\begin{figure}
\begin{center}
\includegraphics*[width=9cm]{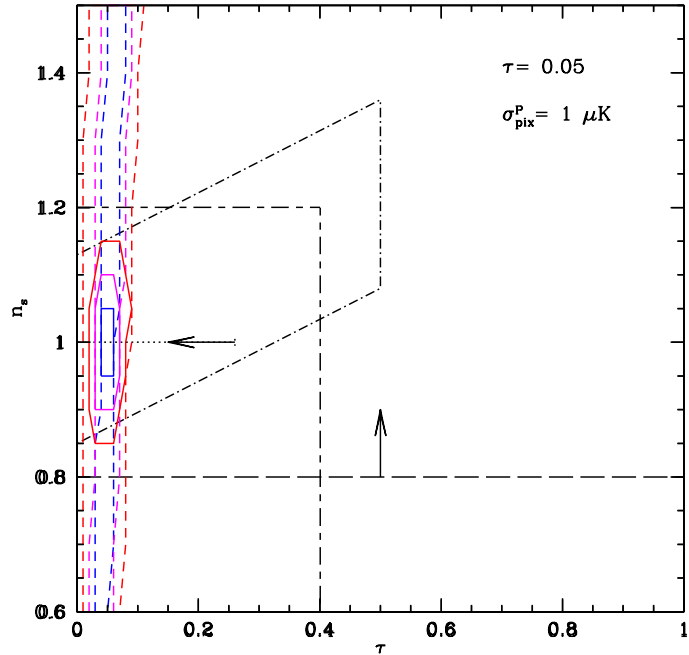}
\end{center}
\caption{Average likelihood distributions, in a model with 
$\tau = 0.05$ and $\sigma^P_{pix} = 1\, \mu$K, if the pixel--pixel 
angular distance is 3.7$^o$. Solid and dashed curves are as in 
Fig.~3. This plot shows that, at this noise level, $\tau$ could 
be fairly reliably detected.}
\label{6fig}
\end{figure}

\begin{figure}
\begin{center}
\includegraphics*[width=9cm]{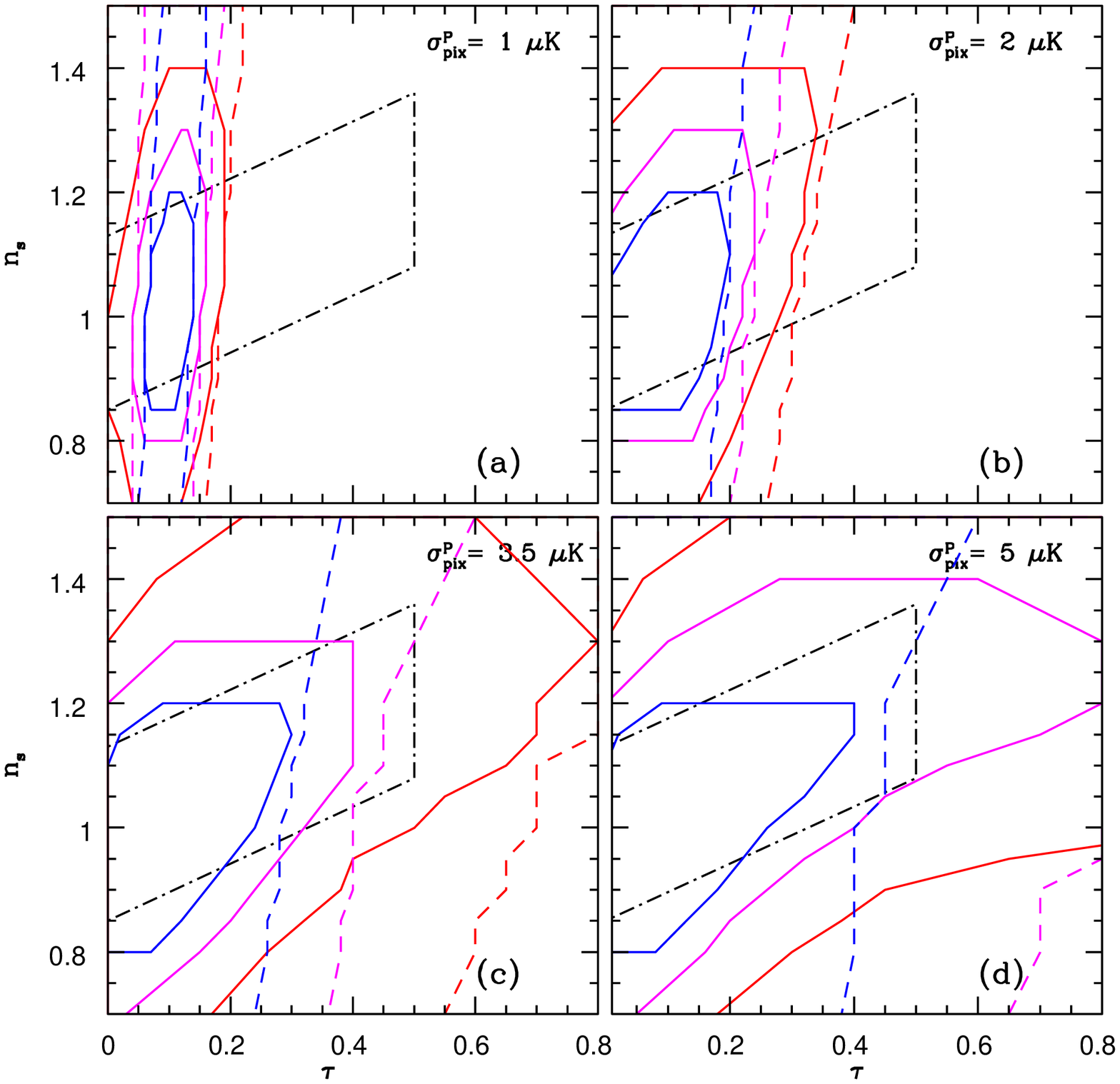}
\end{center}
\caption{Average likelihood distributions, in a model with 
$\tau = 0.10$, if the pixel--pixel 
angular distance is 7.3$^o$. Solid and dashed curves are as in Fig.~3.}
\label{7fig}
\end{figure}

\begin{figure}
\begin{center}
\includegraphics*[width=9cm]{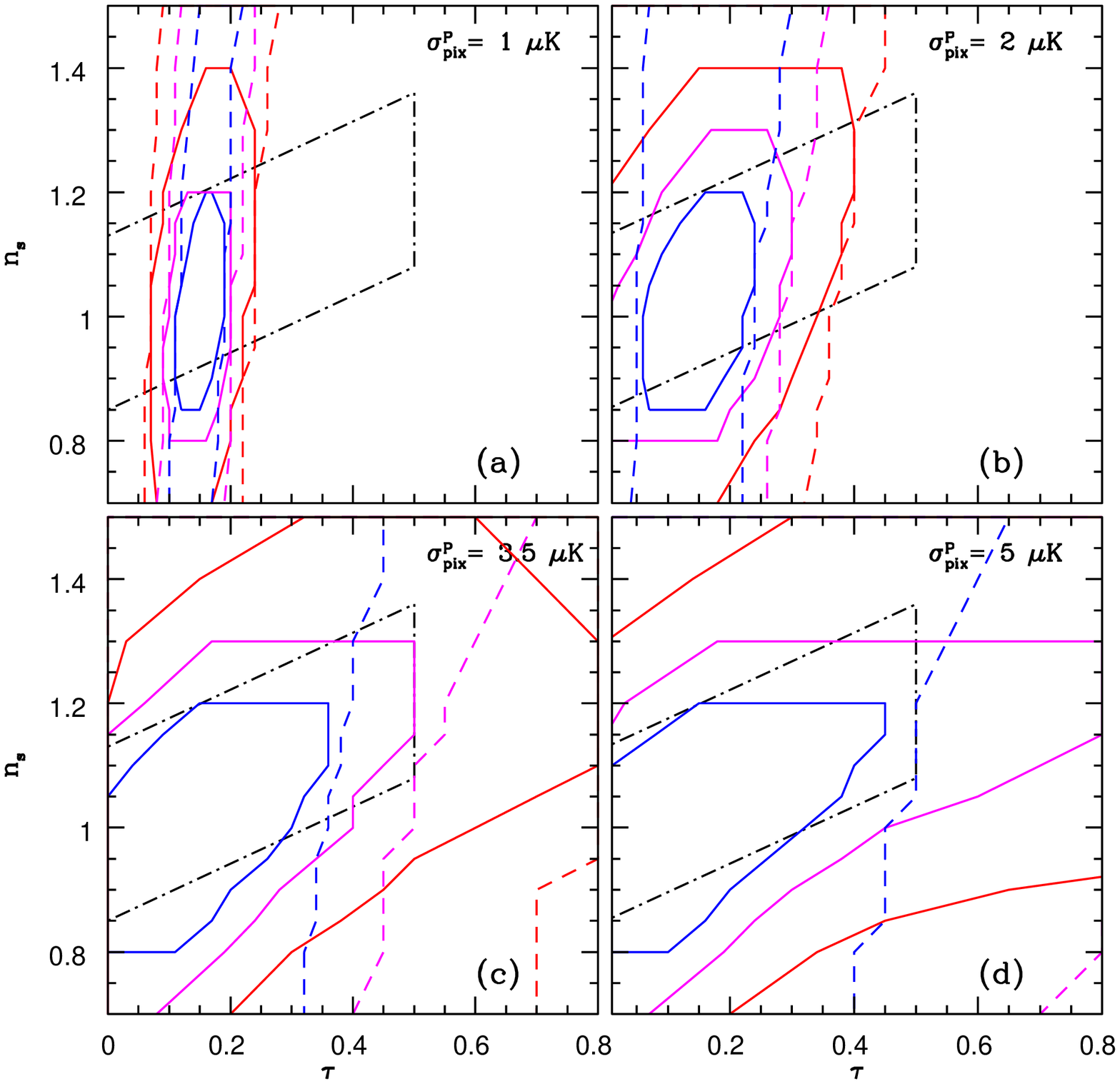}
\end{center}
\caption{Average likelihood distributions, in a model with 
$\tau = 0.15$, if the pixel--pixel 
angular distance is 7.3$^o$. Solid and dashed curves are as in Fig.~3.}
\label{8fig}
\end{figure}

\begin{figure}
\begin{center}
\includegraphics*[width=9cm]{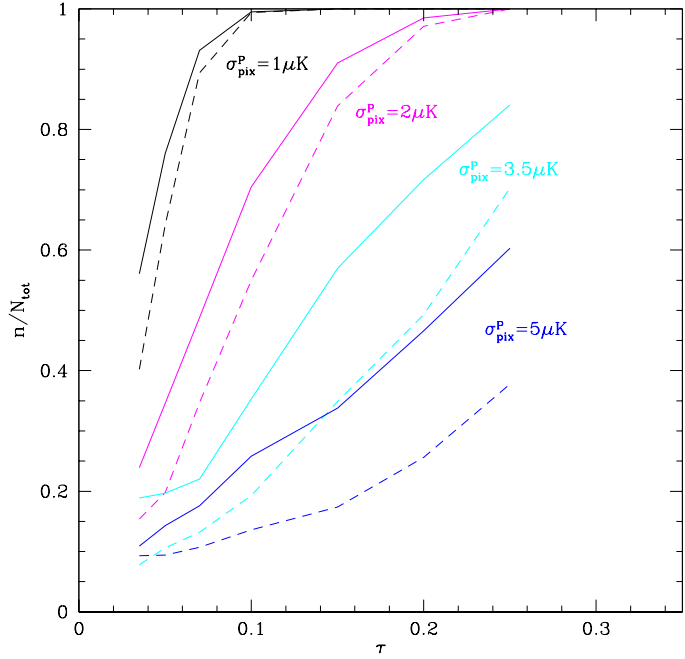}
\end{center}
\caption{Fractions of realizations for which lower limits on $\tau$ 
are detected, at the 1--$\sigma$ confidence level, for different 
$\tau$ and $\sigma^P_{pix}$ values. Solid lines are obtained with 
anisotropy and polarization data. Dashed lines are obtained just with 
polarization. The values of $\sigma^P_{pix}$ reported in the plot, 
refer to a coarse pixelization analysis. For low values of $n/N_{tot}$, 
random noise causes some irregular behaviour.}
\label{9fig}
\end{figure}

\begin{figure}
\begin{center}
\includegraphics*[width=9cm]{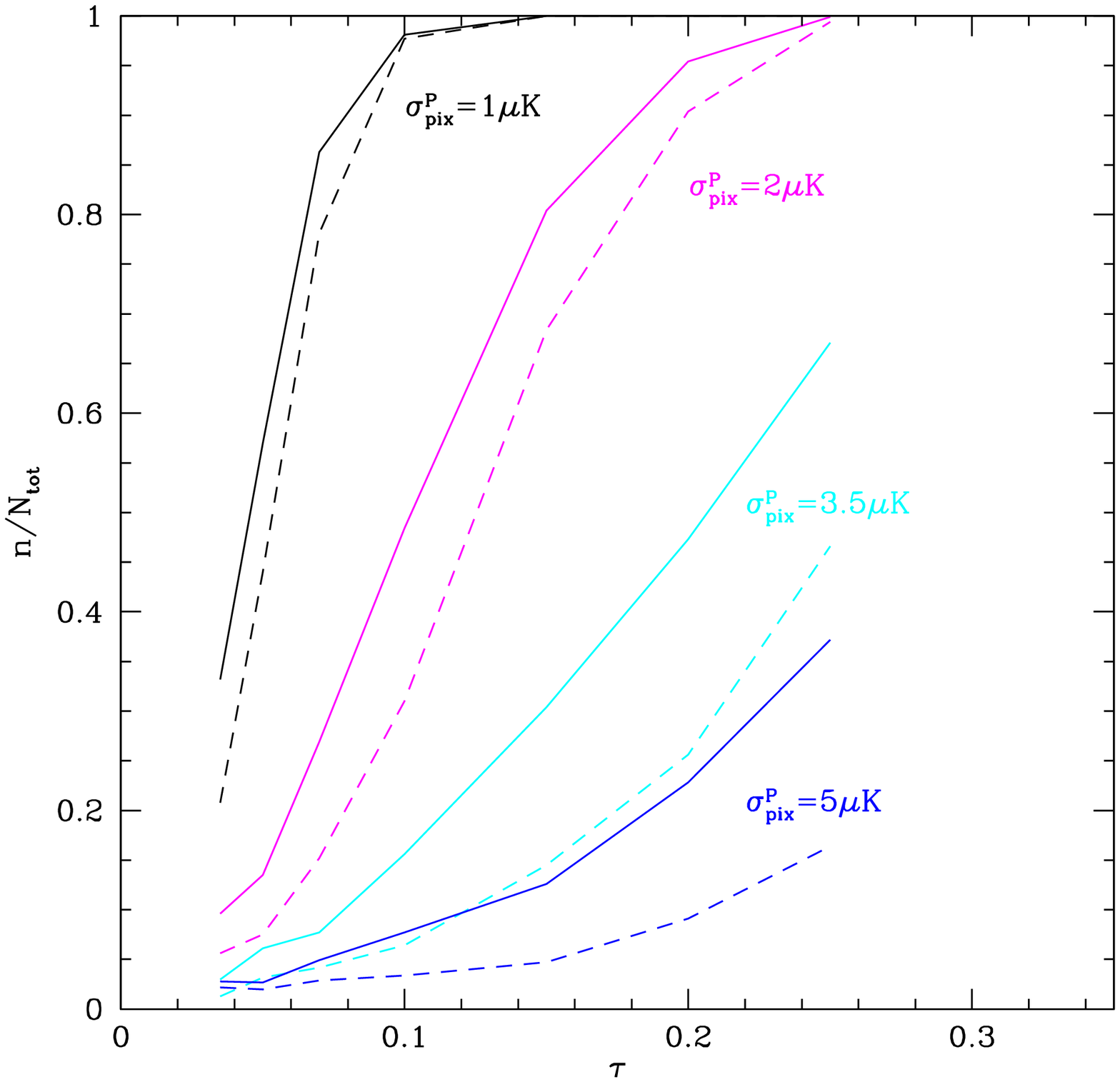}
\end{center}
\caption{As Fig.~9, at the 2--$\sigma$ confidence level.}
\label{10fig}
\end{figure}

\begin{figure}
\begin{center}
\includegraphics*[width=9cm]{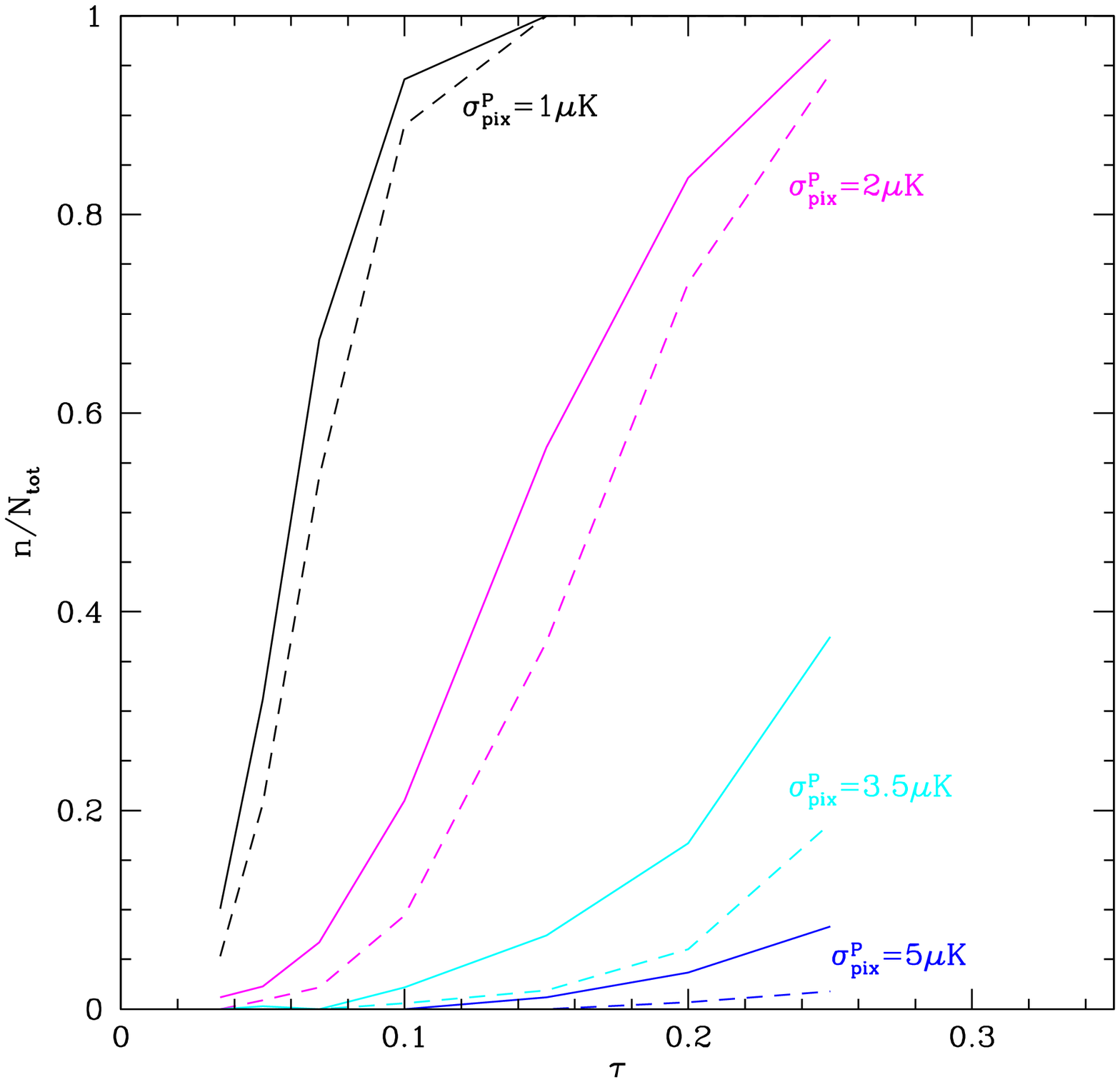}
\end{center}
\caption{As Fig.~9, at the 3--$\sigma$ confidence level.}
\label{11fig}
\end{figure}

\end{document}